\documentclass[twocolumn,amsmath,aps,fleqn]{revtex4}
\usepackage{graphicx,amssymb}
\begin{document}
%My commands
\newcommand{\be}{\begin{equation}}
\newcommand{\ee}{\end{equation}}
\newcommand{\bq}{\begin{eqnarray}}
\newcommand{\eq}{\end{eqnarray}}
\newcommand{\bsq}{\begin{subequations}}
\newcommand{\esq}{\end{subequations}}
\newcommand{\bc}{\begin{center}}
\newcommand{\ec}{\end{center}}
\newcommand {\R}{{\mathcal R}}
\newcommand{\al}{\alpha}
\newcommand\lsim{\mathrel{\rlap{\lower4pt\hbox{\hskip1pt$\sim$}}
    \raise1pt\hbox{$<$}}}
\newcommand\gsim{\mathrel{\rlap{\lower4pt\hbox{\hskip1pt$\sim$}}
    \raise1pt\hbox{$>$}}}

\title{Constraining Logotropic Unified Dark Energy Models}

\author{V. M. C. Ferreira}
\email[Electronic address: ]{vasco.ferreira@astro.up.pt}
\affiliation{Instituto de Astrof\'{\i}sica e Ci\^encias do Espa{\c c}o, Universidade do Porto, CAUP, Rua das Estrelas, PT4150-762 Porto, Portugal}
\affiliation{Centro de Astrof\'{\i}sica da Universidade do Porto, Rua das Estrelas, PT4150-762 Porto, Portugal}
\affiliation{Departamento de F\'{\i}sica e Astronomia, Faculdade de Ci\^encias, Universidade do Porto, Rua do Campo Alegre 687, PT4169-007 Porto, Portugal}

\author{P. P. Avelino}
\email[Electronic address: ]{pedro.avelino@astro.up.pt}
\affiliation{Instituto de Astrof\'{\i}sica e Ci\^encias do Espa{\c c}o, Universidade do Porto, CAUP, Rua das Estrelas, PT4150-762 Porto, Portugal}
\affiliation{Centro de Astrof\'{\i}sica da Universidade do Porto, Rua das Estrelas, PT4150-762 Porto, Portugal}
\affiliation{Departamento de F\'{\i}sica e Astronomia, Faculdade de Ci\^encias, Universidade do Porto, Rua do Campo Alegre 687, PT4169-007 Porto, Portugal}

\date{\today}
\begin{abstract}

A unification of dark matter and dark energy in terms of a logotropic perfect dark fluid has recently been proposed, where deviations with respect to the standard $\Lambda {\rm CDM}$ model are dependent on a single parameter $B$. In this paper we show that the requirement that the linear growth of cosmic structures on comoving scales larger than $8 h^{-1} \, {\rm Mpc}$ is not significantly affected with respect to the standard $\Lambda {\rm CDM}$ result provides the 
strongest constraint to date on the model ($B <6 \times 10^{-7}$), an improvement of more than three orders of magnitude over previous constraints on the value of $B$. We further show that this constraint rules out the logotropic Unified Dark Energy model as a possible solution to the small scale problems of the $\Lambda$CDM model, including the cusp problem of Dark Matter halos or the missing satellite problem, as well as the original version of the model where the Planck energy density was taken as one of the two parameters characterizing the logotropic dark fluid.

\end{abstract}
%\pacs{}
%\keywords{Cosmology; Dark energy}
\maketitle 

\section{\label{intr}Introduction}

Over the past years increasingly precise cosmological observations have been providing solid evidence in favour of the standard cosmological paradigm, the so-called $\Lambda$CDM model (see, e.g., \cite{Suzuki:2011hu,Anderson:2012sa,Ade:2015xua}). This model relies on a cosmological constant to drive the current acceleration of the Universe and on the existence of a Cold Dark Matter (CDM) component to explain the observed properties of the large scale structures of the Universe. The standard cosmological paradigm is also based on the assumptions that the universe is homogeneous and isotropic on cosmological scales, and that General Relativity (GR) accurately describes gravitational interactions at the classical level.

Despite its simplicity and some problems in describing non-linear structures on small scales \cite{Klypin:1999uc,deBlok:2009sp}, the $\Lambda$CDM model is in extremely good agreement with the available cosmological observational data \cite{Ade:2015xua}. Still, there is presently no satisfactory explanation for the tiny energy density associated to the cosmological constant \cite{Carroll:2000fy}. Hence, dynamical Dark Energy (DE) and modified gravity models need to be taken into consideration in the search for a fundamental explanation for the acceleration of the Universe, both at early and late cosmological times \cite{Copeland:2006wr,Frieman:2008sn,Caldwell:2009ix,Li:2011sd,Avelino:2016lpj}.

An interesting class of dynamical DE models, dubbed Unified Dark Energy (UDE) models, considers DE and DM as different manifestations of the same underlying fluid, often assumed to be a perfect fluid (see, e.g., \cite{Beca:2005gc,Avelino:2014nva,Avelino:2015dwa}). Regardless of the particular  parameterizations of the Equation of state (EoS) parameter, the UDE fluid behaves as DM at early times and as DE at late times, and typically has a relatively large sound speed at late times.

The logotropic UDE model \cite{Chavanis:2015eka,Chavanis:2015paa} has recently been proposed as a viable alternative to the $\Lambda$CDM model. In particular, it has been suggested that the small scale problems of the $\Lambda$CDM model, including the missing satellite problem \cite{Klypin:1999uc} and the cuspy halo problem \cite{deBlok:2009sp}, would be absent in the logotropic UDE model. Here, we shall re-examine this proposal, determining the most stringent constraint to date on the logotropic UDE model and discussing the implications of our results to its suggested role as a viable way out to the small scale problems of the standard cosmological model.

Throughout this paper we use units such that $c=1$, where $c$ is the value of the speed of light in vacuum.

\section{Logotropic UDE fluid\label{sec2}}

Let us start by considering a perfect dark fluid model with an EoS of the form $p=p\left(n\right)$, where $p$ is the pressure and $n$ is the particle number density in the comoving free falling frame. Assuming adiabaticity and conservation of the particle number $N$, one may express the local form of the first law of thermodynamics as
\begin{equation}\label{first law}
dU+pdV\propto d\left(\frac{\rho}{n}\right)+pd\left(\frac{1}{n}\right)=0,
\end{equation}
where $U=\rho V$ is the internal energy and $V=N/n$ is the physical volume.

Integrating Eq. \eqref{first law} one obtains
\begin{equation}\label{energy density model II}
\begin{aligned}
\frac{\rho}{n} & =\int^{n} p\left(n'\right)\frac{dn'}{n'^{2}}+m \Leftrightarrow \\
& \Leftrightarrow \rho = nm+n\int^{n}\frac{p\left(n '\right)}{n'^{2}}dn ' = n+u\left(n\right).
\end{aligned}
\end{equation}
The integration constant is such that $\rho=nm$ if $p=0$, where $m$ may be identified as the rest mass of the particles, that we shall take as our unit of mass (i.e., $m=1$). The first term $n$ is characteristic of matter (i.e. a pressureless fluid) while  the second term $u(n)$ shall be used to model DE effects. 

Let us assume that the universe is composed by a single UDE fluid where the logotropic EoS is given by
\begin{equation}\label{logotropic EoS II}
p=A\ln\left(\frac{n}{n_{*}}\right)\,,
\end{equation}
$n_{*}$ being a reference number density. By appropriately choosing $A$ and $n_{*}$ the pressure can be tuned  to be as close to any given constant as one desires, thus implying that the parameters of the model can be adjusted so that it becomes as close to the $\Lambda {\rm CDM}$ model as intended. Equations \eqref{energy density model II} and \eqref{logotropic EoS II} imply that the energy density is given by
\begin{equation}\label{logotropic energy density}
\rho = n -A\ln\left(\frac{n}{n_{*}}\right)-A\,.
\end{equation}
The minimum energy density 
\be
\rho_{min}=-A \ln \left(\frac{A}{n_*}\right)
\ee
is attained for $n=A > 0$ (note that if $n=A$ then $d \rho/dn=0$, $d^2 \rho/dn^2=A/n^2>0$ and $p=-\rho$). 

The relation between the pressure $p$ and the energy density $\rho$ may be written as
\begin{equation}\label{pressure_implicit}
\rho = n_{*}\exp\left(\frac{p}{A}\right)-p-A\,,
\end{equation}
or, alternatively, as
\begin{equation}\label{pressure_fW}
p=-A W_{m}\left[f(\rho)\right]-\rho-A\,,
\end{equation}
where 
\begin{equation}
f(\rho)=-\frac{n_{*}}{A}\exp\left(-\frac{\rho}{A}-1\right)\,,
\end{equation}
$W_{m}$ is the Lambert W function and the subscript $m=0,-1$ refers to the two real branches of this multivalued function (note that $f < 0$ has a minimum equal to $-1/e$ for $\rho=\rho_{min}$ and that $W_m(-1/e)=-1$). In the relevant regime, $-1/e\leq x  <0$, the Lambert W function has two possible branches, $W_{0}\left(x\right) > -1$ (usually called the principal branch) and $W_{-1}\left(x\right) < -1$. Hence, there is a clear physical interpretation of the two real branches of the W Lambert function. The branch $W_{-1}$ corresponds to a standard evolution of the energy density ($d\rho/dn>0$) and $W_{0}$ to a phantom regime ($d\rho/dn<0$). 

The sound speed squared is given by
\begin{equation}\label{eq_sound_speed}
c_{s}^{2} \equiv \frac{dp}{d\rho}=-\left[1+f(\rho)\exp\left(-W_{m}\left[f(\rho)\right]\right)\right]^{-1}\,,
\end{equation}
which is positive in the $W_{-1}$ branch, negative in the $W_{0}$ branch and divergent at $\rho=\rho_{min}$ ($f(\rho_{min})=-1/e$). 

In the matter era $n \sim \rho$ and, consequently,
\begin{equation}\label{logotropic EoS IIM}
p = A\ln\left(\frac{\rho}{n_{*}}\right)\,,
\end{equation}
and
\begin{equation}\label{eq_sound_speedM}
c_{s}^{2} = \frac{A}{\rho}\,,
\end{equation}
to a good approximation.

\section{Logotropic Cosmology\label{sec2}}

In this section we shall consider a flat homogeneous and isotropic universe made up entirely of a logotropic perfect dark fluid of density $\rho$ and pressure $p$. Energy-momentum conservation implies that
\be\label{continuity_eq}
{\dot \rho}+3H\left(\rho+p\right)=0\,,
\ee
where $H=\dot{a}/a$ is the Hubble function, $a$ is the scale factor normalized to unit at the present time (i.e. $a_{0}=1$), and a dot represent a derivative with respect to the physical time. 

Equation (\ref{logotropic energy density}) may be written as
\be
\rho = \rho_{m} + \rho_{de}\,,
\ee
where
\be
\rho_{m}=n  \label{rhom}
\ee
and
\be
\rho_{de}=-A\ln\left(\frac{n}{n_{*}}\right)-A \label{rhode}
\ee
are identified as the matter and DE components, respectively.  Taking into account that $n=n_0 a^{-3}$, it is straightforward to show that
\bq
\frac{\rho_{m}}{\rho_0}&=&\frac{\rho_{m0}}{\rho_0}a^{-3}=\Omega_{m0} a^{-3}\,, \label{rhomovrho0}\\
\frac{\rho_{de}}{\rho_0}&=&\frac{\rho_{de0}}{\rho_0} +3 \frac{A}{\rho_0} \ln a=\Omega_{de0}+3\frac{A}{\rho_0} \ln a \label{rhodeovrho0}\,,
\eq
where the subscript `$0$' again refers to the present time.

The Friedmann equation 
\begin{equation}\label{Friedmann eq}
H^{2}=\frac{8\pi G}{3}\rho 
\end{equation}
may therefore be written as
\begin{equation}\label{Friedmann eq1}
H^2=H_0^2 \left[\Omega_{m0} a^{-3}+\Omega_{de0}\left(1+3 B \ln a \right)\right]\,,
\end{equation}
where
\be
\label{B}
B =\frac{A}{\rho_{de0}}\,.
\ee
Hence, in the $B \to 0$ limit the logotropic UDE model approaches the $\Lambda$CDM model.

The value of the scale factor at the transition between the normal and the phantom regime (when $\rho=\rho_{min}$) is given by
\be
a(\rho=\rho_{min})=\left(\frac{\Omega_{m0}}{B\Omega_{de0}}\right)^{1/3}\,.
\ee
For $B < \Omega_{m0} / \Omega_{de0}$ this transition only happens in the future, and, as expected, it never occurs in the $B \to 0$ limit. 

Although baryons have not explicitly been taken into account in the present section, their inclusion in the definition of the matter density is straightforward.

\section{Observational constraints: growth of cosmic structures\label{sec3}}

Since the model admits a non-zero sound speed, the growth of linear density perturbations can only occur on scales larger than the comoving Jeans length, while on smaller scales pressure gradients give rise to acoustic oscillations. Deep into the matter era, the comoving Jeans length is given approximately by
\begin{equation}
\lambda_{J}^{c}=c_{s}\left(1+z\right)\left(\frac{\pi}{G\rho}\right)^{1/2}\,.
\end{equation}
Using Eqs. (\ref{eq_sound_speedM}) and (\ref{B}), taking into account that $\rho_{de0} = 3 H_0^2 \, \Omega_{de0}/(8\pi G)$ and that, deep into the matter era, $\rho \sim 3 H_0^2 \, \Omega_{m0} \, (1+z)^3/(8\pi G)$, one obtains
\begin{equation}
\lambda_{J}^{c}=\left( \frac{8\pi^2 B}{3}\right)^{1/2} H_0^{-1} \, \Omega_{m0}^{-1} \,  \Omega_{de0}^{1/2} \, (1+z)^{-2}\,.
\end{equation}
By imposing that $\lambda_{J}^{c} < R$ one obtains the following constraint on $B$
\begin{equation}
\label{Bcons}
B<\frac{3}{8\pi^2}(RH_0)^2 \, \Omega_{m0}^2 \, \Omega_{de0}^{-1} \, (1+z)^{4}\,.
\end{equation}
If one requires that the linear growth of cosmic structures on comoving scales larger than $8 h^{-1} \, {\rm Mpc}$ is not significantly affected with respect to the standard matter era evolution for $z>1$ or, equivalently, by taking $z=1$, $R=8 h^{-1} \rm {\rm Mpc}$, $H_0^{-1}=3 \times 10^3 h^{-1} \, {\rm Mpc}$ in Eq. (\ref{Bcons}) together with the latest Planck constraints on the values of the cosmological parameters \cite{Ade:2015xua} ($\Omega_{m0}=0.308$ with $\Omega_{de0}=1-\Omega_{m0}$, $h=0.678$), one finally obtains
\begin{equation}
\label{Bcons1}
B< 6 \times 10^{-7}\,.
\end{equation}
This constraint on the value of $B$ improves the constraints presented in \cite{Chavanis:2015eka,Chavanis:2015paa} by more than three orders of magnitude. Also, the upper limit on the value of $B$ is much smaller than the value ($B \sim 3.5 \times 10^{-3}$) required for the model to play a role as a possible solution to the cusp problem of DM halos or the missing satellite problem.

\section{Role of the Planck density in the logotropic UDE model\label{sec4}}

The EoS of the logotropic dark fluid has two fundamental parameters, $A$ and $n_*$. The parameter $A$ is equal to the minimum energy density of the model which, unless the parameter $B$ is extremely small, should not differ by many orders of magnitude from the observationally inferred present value of the dark energy density $(A = B \rho_{de0})$. On the other hand $n_*$, with
\be
\frac{n_*}{\rho_{de0}} = \frac{\Omega_{m0}}{\Omega_{de0}} \exp\left(1+ \frac{1}{B}\right) \label{nstarovrhode0}\,.
\ee
is a priori a free energy density parameter which, given $\rho_{de0}$ and $\Omega_{m0}$, is a function of the dimensionless parameter $B$ alone (we have used Eqs. (\ref{rhom}), (\ref{rhode}), (\ref{rhomovrho0}) (\ref{rhodeovrho0})  and (\ref{B}) in the derivation of Eq. (\ref{nstarovrhode0})). 

In \cite{Chavanis:2015eka,Chavanis:2015paa} $n_*$ has been identified with the Planck density $\rho_{pl}$. With this identification in place, $B$ is no longer a free parameter:
\be
B=B_{pl}=\left(\ln \left(\frac{\Omega_{de0}}{\Omega_{m0}} \frac{\rho_{pl}}{\rho_{de0}}\right) -1\right)^{-1} = 3.5 \times 10^{-3}\,.
\ee
However, as shown in the previous section, $B$ must be smaller than $6 \times 10^{-7}$ in order that the linear growth of cosmic structures on comoving scales larger than $8 h^{-1} \, {\rm Mpc}$ is not significantly affected with respect to the standard $\Lambda$CDM result, thus implying that a value of $B$ equal to $B_{pl}$ is simply ruled out. 

In fact the $B < 6 \times 10^{-7}$ constraint implies that
\be
\log_{10} \left(\frac{n_*}{\rho_{pl}}\right) = \log_{10} \left(\exp\left(\frac{1}{B}-\frac{1}{B_{pl}}\right)\right) >  7 \times 10^5\,,
\ee
thus leading to a value of $n_*$ which is many orders of magnitude larger than the Planck density. Such a large value of the fundamental density of the logotropic fluid model would be hard to justify at a more fundamental level.

\section{Small scale problems of the $\Lambda$CDM model \label{sec5}}

In the non-relativistic limit, with $n \sim \rho$, $c_s^2 \equiv dp/d\rho=A/\rho \ll 1$ and $|w|=|p/\rho| \ll 1$, a logotropic sphere in hydrostatic equilibrium satisfies
\be
\frac{1}{\rho}\frac{dp}{dr}=\frac{A}{\rho^2} \frac{d\rho}{dr}=-\frac{GM}{r^2}\,.\label{hydrostatic}
\ee 
Substituting the ansatz
\be
M(r) \propto \rho r^3\,, \qquad \frac{d \rho}{dr} \propto \frac{\rho}{r}
\ee
into Eq. (\ref{hydrostatic}), one finds that the surface density of the logotropic sphere is a constant proportional to $B^{1/2}$,
\be
\rho r \propto \left(\frac{A}{G}\right)^{1/2} =  \left(\frac{B \rho_{de0}}{G}\right)^{1/2}\,. \label{surfden}
\ee
In \cite{Chavanis:2015eka,Chavanis:2015paa} it was claimed that a constant surface density of DM halos consistent with observations  \cite{Donato:2009ab} (see also \cite{Saburova:2014opa} for a different perspective) would be obtained if $n_* = \rho_{pl}$, or equivalently, $B=3.5 \times 10^{-3}$. Of course, for $B =  6 \times 10^{-7}$ the estimate of the surface density of DM halos would decrease by a factor of almost two orders of magnitude, and the claimed agreement with observations would be lost. The same also applies to the claim that the logotropic equation of state with $n_* = \rho_{pl}$ would also explain the observed Tully-Fisher relation and the mass of dwarf galaxies, since both of these depend strongly on the surface density of the DM halos.

Also note that the equilibrium solution presented in Eq. (\ref{surfden}) is unstable, even far away from the singularity at $r=0$. This may be shown by considering the lagrangian perturbation $r \to \lambda r$, leading to $\rho \to \lambda^{-3} \rho$, where $\lambda$ is a constant smaller than unity (the sphere is contracted). In this case
\bq
\frac{1}{\rho}\frac{dp}{dr} = \frac{A}{\rho^2} \frac{d\rho}{dr} &\to& \lambda^2 \frac{1}{\rho}\frac{dp}{dr}\,, \\
-\frac{GM}{r^2} &\to& - \lambda^{-2} \frac{GM}{r^2} \,.
\eq
Hence, such a perturbation would lead to an increase of the acceleration due to gravity and a decrease of the acceleration due to the pressure gradients, thus implying that the pressure gradients would not be able to stop the logotropic sphere from further collapse. This shows that the above hydrostatic logotropic equilibrium solution is unstable. 

If $n_* = \rho_{pl}$ ($B=B_{pl}=3.5 \times 10^{-3}$) then the comoving Jeans length at the start of the matter era is of the same order of magnitude of the smallest known dark matter halos. In \cite{Chavanis:2015eka,Chavanis:2015paa} this has been suggested as a possible explanation to the minimum observed size of DM halos in the universe. It was further suggested that this suppression of small scale power could be a way to solve the missing satellite problem in the context of the logotropic UDE model. Again, for $B = 6 \times 10^{-7}$ the sound speed is smaller by almost two orders of magnitude than $B_{pl}$ and, therefore, the estimate of the minimum size of dark matter halos would be smaller by the same amount. This precludes the logotropic model from providing an explanation for the minimum size of DM halos or a solution to the missing satellite problem.

\section{\label{conc} Conclusions}

In this paper we derived the most stringent constraint to date on the recently proposed logotropic UDE model, improving previous constraints on the parameter $B$ by more than three orders of magnitude. We have shown that this constraint rules out the original version of the logotropic UDE model, showing that the characteristic density $n_*$ would have to be more than $7 \times 10^5$ orders of magnitude larger than the Planck density in order for the model to be consistent with observations. Furthermore, we have found that the sound speed of the logotropic dark fluid is not large enough for it to constitute a viable  alternative to the $\Lambda$CDM model in the search for a solution to its problems in describing the structures on small non-linear scales, including  the cusp problem of Dark Matter halos or the missing satellite problem. Still, although the main motivation for the logotropic UDE model has been lost, if the parameter $B$ is small enough it remains an interesting DE parameterization which may be relevant when considering small deviations with respect to the $\Lambda$CDM model consistent with the current cosmological observational data.

%%%%%%%%%%%%%%%%%%%%%%%%%%%%%%%%%%%%%%%%%%%%%%%%%%%%%
\begin{acknowledgments}

The authors are grateful to Alberto Rozas-Fernandez and to the other members of the Cosmology Thematic Line of the Institute of Astrophysics and Space Sciences for enlightening discussions. V.M.C. Ferreira was supported by FCT through national funds and by FEDER through COMPETE2020 (ref: POCI-01-0145-FEDER-007672).  P.P.A. was supported by FCT through the Investigador FCT contract reference IF/00863/2012 and POPH/FSE (EC) by FEDER funding through the program Programa Operacional de Factores de Competitividade - COMPETE. Funding of this work was also provided by the FCT grant UID/FIS/04434/2013.

\end{acknowledgments}

%%%%%%%%%%%%%%%%%%%%%%%%%%%%%%%%%%%%%%%%%%%%%%%%%%%%%%%%%%

\bibliography{logotropic}

\end{document}